\documentclass[twocolumn,prd,aps,amsmath,amssymb,epsfig]{revtex4}

\usepackage{graphicx}
\usepackage{dcolumn}
\usepackage{bm}

\begin{document}
\title{${\cal N}=1$ Super Yang-Mills from Supergravity: The UV-IR
Connection}

\author{Poul {\sc Olesen}}
\email{polesen@nbi.dk}
\author{Francesco {\sc Sannino}}%
 \email{francesco.sannino@nbi.dk}
\affiliation{The Niels Bohr Institute \& {\rm NORDITA},
Blegdamsvej 17, DK-2100 Copenhagen \O, Denmark}


\begin{abstract}

We consider the Maldacena-Nu$\tilde{\rm n}$ez supergravity
solution corresponding to ${\cal N}=1$ super Yang-Mills within the
approach by Di Vecchia, Lerda and Merlatti. We show that if one
uses the radial distance as a field theory scale, the
corresponding beta function has an infrared fixed point. Assuming
this to be a physical property for all four dimensional
non-singular renormalization schemes, we use the relation between
the gaugino condensate and its dual to investigate the connection
between the IR and UV behaviors. Imposing the ``field theory
boundary condition'' that the first two terms in the perturbative
UV beta function are universal, the fixed point is found to be of
first order, and the slope of the IR beta function is also fixed.
\end{abstract}

\maketitle

\section{Introduction}
\label{introduction}

 Maldacena and ${\rm Nu\tilde{n}ez}$
\cite{Maldacena:2000yy} found a geometry which is dual to a little
string theory that reduces to pure ${\cal N}=1$ super Yang-Mills
in the IR. The MN gravity solution corresponds to a large number
of NS or D fivebranes wrapped on a two sphere. This supergravity
solution is regular and breaks the $U(1)_R$ chiral symmetry in the
expected manner. Recently the authors in
\cite{{Apreda:2001qb},DiVecchia:2002ks} discussed four-dimensional
${\cal N}=1$ and ${\cal N}=2$ super Yang-Mills theories with gauge
group $SU(N)$. In the ${\cal N}=1$ case an interesting connection
between the geometric radial variable and the field theory scale
was obtained by use of the gravitational dual of the gaugino
condensate. These authors considered in details the supergravity
solutions corresponding to $N$ D5-branes wrapped on a two-sphere
that were first found in
\cite{Maldacena:2000yy,{Apreda:2001qb},{DiVecchia:2002ks},{JPG},FB}
and they then considered the massless open-string modes
propagating on the flat part of the D5 world-volume. An effective
action at energies where the higher string modes and the
Kaluza-Klein excitations around the two-cycles decouple was found.
The resulting theory is four-dimensional $SU(N)$ super Yang-Mills
theory. In the following we consider only the most interesting
case ${\cal N}=1$. The MN supergravity solution is dual to a gauge
theory with four supercharges in four dimensions. In the $SO(4)$
gauged supergravity Lagrangian one of the two $SU(2)$ factors
($SU(2)_L$) inside $SO(4)$ is gauged by imposing an anti-self
duality constraint. This field has a Yang-Mills coupling $\lambda$
which is dimensionfull. The inverse square of the four-dimensional
coupling is proportional to the volume of the two-sphere on which
the D5 branes are wrapped. It then turns out that the coupling
constant $g_{\rm YM}$ and the vacuum angle $\theta_{\rm YM}$ can
be expressed in terms of the ten-dimensional supergravity solution
representing the $N$ wrapped D5 branes in an explicit form,
\begin{equation}
\frac{1}{g_{\rm YM}^2}=G(\rho), \label{1}
\end{equation}
found by Di Vecchia, Lerda and Merlatti \cite{DiVecchia:2002ks}.
Here $\rho$ is a dimensionless radial variable the scale of which
is given in terms of the dimensionfull radial variable $r$ and the
coupling $\lambda$, $\rho=\lambda r$, and $G$ is a known
function. To obtain this result a Born-Infeld analysis was used.

In general the radial variable $\rho$ should be considered as a
function of the four-dimensional field theory scale $\mu/\Lambda$,
where $\mu$ is the variable scale and $\Lambda$ is a
renormalization group invariant scale. The aim of this work is to
investigate in some detail such a relation. In Reference
\cite{DiVecchia:2002ks} this connection was deduced by relating
the MN solution to the non-vanishing of the gaugino condensate
\cite{Apreda:2001qb} in ${\cal N}=1$ super Yang-Mills without
matter. As we shall see there is a considerable amount of freedom
in the choice of the relation, and some guiding principle is
needed. The relation (\ref{1}) is, however, free of any such
ambiguity, and hence we first compute the following beta-function
\begin{equation}
\beta_\rho (g_{\rm YM})=\frac{\partial g_{\rm YM}}{\partial\rho}.
\label{2}
\end{equation}
Since $\rho=\rho (\mu/\Lambda)$, this is a well defined
beta-function, which can be computed directly. As our guiding
principle we then {\it assume} that any ``physical'' result which
can be derived from $\beta_\rho$, should hold also for
\begin{equation}
\beta (g_{\rm YM})=\frac{\partial g_{\rm YM}}{\partial\ln\mu
/\Lambda}. \label{3}
\end{equation}
It is well known that in general this $\beta$ is rather arbitrary
since it is scheme dependent. However, asymptotic freedom and a
possible IR fixed point in $\beta$ are expected to be scheme
independent and as such to be relevant ``physical'' information.

Now it turns out that $\beta_\rho$ possesses an IR fixed point. In
going from $\beta_\rho$ to $\beta$, this fixed point can be
``removed'' only by an infrared singular transformation. According
to the assumption that an IR fixed point is physical such a
singular transformation is not allowed. {}This is a welcome
feature. Indeed for the correspondence between supergravity and
gauge theories to be meaningful we should not lose the precious
information carried by supergravity computations.

Although the persistence of the IR fixed point in going from
eq.~(\ref{2}) to eq.~(\ref{3}) imposes some constraint on the
relation $\rho =\rho (\mu/\Lambda)$ there is still a large amount
of freedom. The order of the fixed point in $\beta$ is, for
example, not fixed yet. We can resolve this ambiguity by
investigating the situation in the UV region. Here the
asymptotically free first order perturbative beta-function was
shown to follow from the duality relation between the gaugino
condensate and the MN solution introduced in
\cite{DiVecchia:2002ks}. However, from this relation the second
order term in the beta function does not occur with the ``right''
coefficient. It was shown by 't Hooft \cite{Hooft-Book} that if
one restricts to analytic changes in the coupling, the second term
in $\beta$ is universal. Our modification of the relation $\rho
=\rho (\mu/\Lambda )$ used in \cite{DiVecchia:2002ks} can be
adjusted in such a way that the second term in $\beta$ gets the
right coefficient. At the same time this can be used to predict
the IR fixed point to be of first order, and the slope of $\beta$
can also be computed near this fixed point.

We have so shown that if the supergravity--gauge theory relation
is valid the ${\cal N}=1$ Yang-Mills theory possesses in the IR an
infrared stable fixed point. In the UV this theory displays
asymptotic freedom.

 In Section
\ref{Gauge} we briefly summarize some known aspects of the ${\cal
N}=1$ Yang-Mills theory relevant to this paper. In Section
\ref{Wrapped} we first study in some detail the beta function
defined with respect to the supergravity dimensionless radial
coordinate. After having shown that in the supergravity variable
the theory possesses a fixed point we suggest how to introduce a
general relation between the supergravity variable and the four
dimensional renormalization scale $\mu$ still leading to a fixed
point. In Section \ref{link} we constrain the newly introduced
relation by requiring the UV result to match the known two loop
universal coefficient of the super Yang-Mills beta function. This
fixes for us also the behavior in the IR. We then show that super
Yang-Mills possesses a non perturbative IR stable fixed point. We
comment our results when concluding in section \ref{The End}.

\section{Aspects of the Super Yang-Mills
Theory} \label{Gauge}

The $SU(N)$ super Yang-Mills theory can be compactly written as:
\begin{eqnarray}
{\cal L} = \frac{1}{4\pi} {\rm Im} \left[\int d^2\theta \,
\tau_{YM} \, {\rm Tr} \left[W^{\alpha}W_{\alpha}\right] \right]  \
, \label{uno}
\end{eqnarray}
 with
\begin{eqnarray}
\tau_{YM}=\frac{\theta}{2\pi} + i\frac{4\pi}{g^2_{YM}} \ .
\label{due}
\end{eqnarray}
 and $W_{\alpha}=W_{\alpha}^a T^a$ with
$ a=1,\ldots,N^2 - 1$. We normalize our generators according to
\begin{eqnarray}
{\rm Tr}\left[T^a T^b\right] = \frac{1}{2} \delta^{ab} \ , \quad
\left[T^a,T^b\right]=i\, f^{abc}T^c \ . \label{tre}
\end{eqnarray}
At the classical level the theory possesses a $U(1)_R$ symmetry
which does not commute with the supersymmetric algebra. The ABJ
anomaly breaks the $U(1)_R$ symmetry to $Z_{2N}$. Non perturbative
effects trigger the gluino condensation leading to further
breaking of $Z_{2N}$ symmetry to a left over $Z_2$ symmetry and we
have $N$ equivalent vacua. The gluino condensate is a relevant
ingredient to constrain the theory, also at large $N$, see
\cite{Shifman:1999kf},\cite{hollowood} and \cite{davies}. In
the literature it has been
computed in different ways and we recall here its expression in a
much studied scheme \cite{Shifman:1999kf}:
\begin{itemize}
\item{\underline{Pauli Villars Scheme}
\begin{eqnarray}
\langle \lambda^2  \rangle &=& Const.\, \mu^3
Im\left[\frac{\tau_{YM}}{4\pi}\right]e^{ i\frac{2\pi}{N}
{\tau_{YM}}} \nonumber \\ &=& Const.\, \mu^3 \frac{1}{g^2_{YM}} \,
e^{ -\frac{8\pi^2}{N\, g^2_{YM}}} e^{i \frac{\theta}{N}}
\label{gYM}
\end{eqnarray}}
\end{itemize}
The Pauli-Villars scheme leads to an expression for the gluino
condensate which is not holomorphic in $\tau_{YM}$. The gluino
condensate is a universal physical constant independent on the
scheme. This fact allows us to compute the {\it perturbative} beta
function in different schemes since the respective coupling
constant must depend on the scale in such a way to compensate the
dependence on $\mu$.  It also enables us to establish a relation
between the coupling constants among different schemes.

It is possible to define a scheme in which the gluino possesses an
holomorphic dependence in $\tau_{YM}^{H}$ \cite{Shifman:1999kf},
\cite{hollowood}, \cite{davies}.
The holomorphic scheme is very constrained leading to a pure
one-loop type of running. This is a welcome feature from the point
of view of the supercurrent chiral multiplet
\cite{Shifman:1999kf}. The two coupling constants are related in
the following way:
\begin{eqnarray}
{\tau_{YM}^H}=\tau_{YM}-i\frac{N}{2\pi}\ln \left[{\rm Im}\left(
\frac{\tau_{YM}}{4\pi}\right)\right] \ . \label{coupling-relation}
\end{eqnarray}
Note that the two couplings are connected via a non analytical
transformation. In the following we will use the Pauli-Villars
scheme or any other scheme analytically related to it. Some of the
aspects of the schemes non analytically related to the
Pauli-Villars one from the point of view of supergravity will be
investigated in \cite{Marotta-Sannino}.

The independence on the scale of the gluino condensate leads to
the following beta function \cite{Shifman:1999kf}:
\begin{eqnarray}
\beta\left(g_{YM}\right)&=& -\frac{3N}{16\pi^2}\,{g^3_{YM}}\,
\left[{1-\frac{N g_{YM}^2}{8\pi^2}}\right]^{-1} \ .\label{betaym}
\end{eqnarray}
Where $\displaystyle{\beta(g_{YM})=\frac{\partial g_{YM}}{\partial
L}}$ and $\displaystyle{L=\ln (\mu/\Lambda)}$ while $\mu$ is the
renormalization scale and $\Lambda$ a reference scale.

Although this coupling may capture the all order perturbative
terms the non perturbative contributions (if any) to the complete
beta functions are still missing.

It is worth mentioning that the universality argument for the two
loop beta function coefficient is valid for all of the infinite
renormalization schemes whose coupling constants are related by
analytical transformations \cite{Hooft-Book}.  It is also expected
that since the presence of a fixed point carries physical
information as such should be renormalization scheme independent
\cite{Hooft-Book}.

If supergravity-gauge theory correspondence is valid we should be
able to adopt different renormalization schemes when connecting
the supergravity solutions to gauge theories. Besides for this
correspondence to be meaningful we should not lose the precious
information carried in supergravity calculations. We will address
these issues in the forthcoming sections.

\section{Wrapped Branes: The non perturbative fixed point}

\label{Wrapped}

In \cite{DiVecchia:2002ks} the coupling $g_{\rm YM}^2N$  was
studied as a function of the (dimensionless) radial variable
$\rho$. In particular, it was shown that
\begin{equation}\frac{1}{g_{\rm YM}^2N}=\frac{Y(\rho )}{16 \pi^2}E
\left(\sqrt{\frac{Y(\rho )-1}{Y(\rho)}}\right), \label{paolo}
\end{equation} where the function $Y$ is given by \begin{equation} Y(\rho )=4\rho \coth
2\rho -1, \end{equation} and $E$ is a complete elliptic integral
\begin{equation} E(k)=\int_0^{\pi/2}d\phi ~\sqrt{1-k^2\sin^2\phi}. \end{equation} In
deriving this result supergravity was used. In principle, in the
infrared limit $\rho\rightarrow 0$ the superstring (to which
supergravity is an approximation) should be used. However, the
supergravity metric is non-singular for the ${\cal N}=1$
case, in contrast to the ${\cal N}=2$ case.
Therefore it still makes sense to ask for the infrared behavior
even in the supergravity approximation. We therefore expand eq.
(\ref{paolo}) around $\rho=0$, \begin{equation} g_{\rm
YM}\sqrt{N}\approx g_c\sqrt{N}(1-\rho^2+\frac{29}{15}\rho^4+...),
\label{approx} \end{equation} where the critical coupling
$g_c\sqrt{N}=4\sqrt{2\pi}$ is reached for $\rho=0$, which is a
non-singular point in the supergravity metric, as already
mentioned.

{}From eq. (\ref{approx}) it thus seems as if the coupling
``stops'' at some finite value for $\rho\rightarrow 0$. Ultimately
we would like to have a connection between $\rho$ and the field
theory scale $\mu$. Without specifying any explicit relation, let
us therefore consider the function $\rho =\rho (\mu/\Lambda)$,
where $\Lambda$ is a fixed renormalization group invariant scale.
Usually one considers the variation of the coupling with $\mu$ in
terms of the $\beta$ function. However, since $\rho$ is a function
of $\mu$, we could equally well define the beta-function
\begin{eqnarray} \beta_\rho (g_{\rm YM})&\equiv& \frac{\partial g_{\rm
YM}}{\partial\rho} \approx -2g_c\rho \nonumber
\\ &\approx& -2\sqrt{g_c(g_c- g_{\rm YM})},
\label{betarho} \end{eqnarray} where we used eq. (\ref{approx}) in
the last step. {}From eq. (\ref{betarho}) we see that the reason
that the coupling ``stops'' for $\rho\rightarrow 0$ is that the
$\beta_\rho$ function has a zero for $g_{\rm YM}\rightarrow g_c$.
The square root behavior of the zero in $\beta_\rho$ is not expected to occur
in field theory with a conventional scale $\mu$. However, the scale
$\rho$ is not conventional in supersymmetric Yang-Mills theory, and in general
$\rho$ is expected to be a complicated function of $\mu/\Lambda$.

To obtain a conventional beta-function we should use the relation
\begin{equation} \beta (g_{\rm YM})=\beta_\rho (g_{\rm YM})~\frac{\partial
\rho} {\partial L},~~L\equiv \ln \frac{\mu}{\Lambda}.
\label{jacobian} \end{equation} {}From this relation it follows
that the conventional beta-functions also have a zero at $g_{\rm
YM}=g_c$ provided the Jacobian \begin{equation} \frac{\partial
\rho}{\partial L} \label{jac} \end{equation} is {\it non-singular}
when $g_{\rm YM}=g_c$.

It is, of course, possible to choose a singular Jacobian
(\ref{jac}) such that the zero in $\beta_\rho$ is removed by
taking (\ref{jac}) to be proportional to $1/\sqrt{(g_c-g_{\rm
YM})}$. This illustrates the arbitrariness of the beta-function.
However, from the point of view of the supergravity dual this
appears unnatural, since there is clearly a zero in the
$\beta_\rho$ function. {}From the point of view of the field
theory (forgetting about the supergravity) this also appears quite
unnatural, since then the coupling $g_{\rm YM}$ anyhow ``stops'',
but this time for no (good) reason, as we shall see in the
following. Thus, based on the supergravity/super-Yang-Mills
correspondence, it appears natural to assign the zero in the
beta-function a physical value.

In \cite{{Apreda:2001qb},DiVecchia:2002ks} the gravitational dual
of the gaugino condensate was shown to be given by the function
\begin{equation} a(\rho)=\frac{2\rho}{\sinh 2\rho}, \label{a}
\end{equation} which is bounded between 0
($\rho\rightarrow\infty$) and 1 ($\rho\rightarrow 0$). On the
basis of the behavior of the gaugino condensate,
\begin{equation}< \lambda^2>=\Lambda^3={\rm renormalization~group~inv.},
\end{equation} it was then argued in \cite{DiVecchia:2002ks} that the supergravity dual $a(\rho)$
behaves like \begin{equation} a(\rho)=\frac{\Lambda^3}{\mu^3}.
\label{alambda} \end{equation} This fixes the connection between
$\rho$ and $\mu$. However, since $a$ is bounded by 1, it is clear
that one can never reach the infrared limit $\mu\rightarrow 0$,
where $\Lambda/\mu$ becomes infinite. Instead, the lowest possible
scale is $\mu=\Lambda$.

To analyze in more details what happens, let us consider the
behavior of $a$ near $\rho =0$, where it follows from eqs.
(\ref{a}) and (\ref{alambda}) that \begin{equation} e^{-3L}\approx
1-\frac{2}{3}\rho^2+\frac{14}{45}\rho^4 + \ldots, \end{equation}
leading to the Jacobian \begin{equation}
\frac{\partial\rho}{\partial L}\approx
\frac{9}{4\rho}+\frac{3}{5}\rho+\ldots \ . \label{singular}
\end{equation} Thus we see that the connection (\ref{alambda})
leds to a singularity at $\rho =0$, and since from eq.
(\ref{betarho}) $\beta_\rho\propto \rho$, it follows that the zero
has been transformed away by a singular Jacobian. {}For $g_{\rm
YM}\approx g_c$ we can easily compute the beta-function based on
the assumption (\ref{alambda}), \begin{equation} \beta (g_{\rm
YM})\approx g_c~\left(-\frac{9}{2}+\frac{81}{5}\rho^2 +...\right),
\end{equation} which can also be expressed in terms of the coupling through
eq. (\ref{approx}), \begin{equation} \beta (g_{\rm YM})\approx
g_c~\left(-\frac{9}{2}+\frac{81}{5} \frac{(g_c-g_{\rm
YM})}{g_c}+\ldots\right). \label{psresult} \end{equation} Thus we
see that from the field theory point of view it appears as if one
can go to higher coupling than $g_c$, since $\beta$ has no
singularities (poles or cuts) for $g_{\rm YM}>g_c$. {}For somebody
who does not know the supergravity connection, there is no way to
understand why the coupling should come to a stop for
$\mu=\Lambda$. Thus the field theory appears to be intrinsically
mysterious.

The peculiar behavior discussed above is due to the use of eq.
(\ref{alambda}), which implies that $\mu$ can never be smaller
than $\Lambda$ due to $a\leq 1$. Of course, one can argue that
these peculiarities are due to the use of the supergravity
approximation to the superstring, and hence one cannot trust the
infrared behavior. However, as mentioned before, the metric is
regular for $\rho\rightarrow 0$, so there is no obvious reason to
ignore the supergravity approximation. In any case, it is of
interest to see if this approximation can be brought into a
satisfactory context in super-Yang-Mills field theory. To this
end, let us notice that eq. (\ref{alambda}) a priori can be
replaced by \begin{equation} a(\rho )=f(g_{\rm
YM})~\frac{\Lambda^3}{\mu^3}, \label{new} \end{equation} where $f$
is some function of the coupling $g_{\rm YM}$. It should be
emphasized that due to the considerable freedom in the choice of
renormalization schemes, corresponding to different couplings, the
function $f$ is by no means unique.

Let us consider the IR limit $\mu\rightarrow 0$, corresponding to
$\rho\rightarrow 0$. {}From eq. (\ref{new}) we clearly need
$f(g_{\rm YM})\rightarrow 0$ in order that $a\rightarrow 1$.
{}Furthermore, since \begin{equation}
\frac{\mu}{\Lambda}=\exp\left(\int_{g_*}^{g_{\rm YM}(\mu
)}~\frac{dx} {\beta (x)}\right) \end{equation} with $g_*=g_{\rm
YM}(\mu =\Lambda )$, and since $a\rightarrow 1$, we need
\begin{equation} f(g_{\rm
YM})\approx\exp\left(+3\int_{g_*}^{g_{\rm YM}(\mu )}~\frac{dx}
{\beta (x)}\right)\rightarrow 0. \label{IR} \end{equation} This
implies that
\begin{equation} \int_{g_*}^{g_{\rm YM}(\mu )}~\frac{dx}{\beta
(x)}\rightarrow -\infty. \label{nolandau} \end{equation} This is
the condition for having no Landau singularity. In order for this
condition to be satisfied when $g_{\rm YM}\rightarrow g_c$, we
clearly need a zero in the beta function.

The rest of the analysis is standard. Assume the existence of a
zero, \begin{equation} \beta (g_{\rm YM})=-\beta_0 (g_c-g_{\rm
YM})^\alpha, \end{equation} where $\alpha \geq 1$ in order to
satisfy the condition (\ref{nolandau}), we obtain for $\alpha =1$,
\begin{equation} f(g_{\rm YM})\approx {\rm const.}~(g_c-g_{\rm
YM})^\frac{3}{\beta_0}, \end{equation} and for $\alpha >1$ we
obtain
\begin{equation} f(g_{\rm YM})\approx {\rm
const.}~\exp\left(\frac{3}{\beta_0 (1-\alpha)} (g_c-g_{\rm
YM})^{1-\alpha}\right). \end{equation} We also remark that the
standard behavior near the fixed point, i.e. \begin{equation}
g_c-g_{\rm YM}\approx {\rm const.}\,e^{\beta_0 L} \end{equation}
for $\alpha=1$, and
\begin{equation} g_c-g_{\rm YM}\approx (\beta_0
(1-\alpha)L)^{\frac{1} {1-\alpha}}, \end{equation} for $\alpha>1$,
taken together with eq. (\ref{approx}), lead to
\begin{eqnarray}
&&\rho\approx {\rm
const.}~\left(\frac{\mu}{\Lambda}\right)^{\beta_0/2}~~
{\rm for}~~\alpha =1~~{\rm and}\nonumber \\
&&\rho\approx (1/\sqrt{g_c})~(\beta_0 (1-\alpha)
L)^{1/(2(1-\alpha))}~~{\rm for}~~\alpha>1. \label{important}
\end{eqnarray}
These equations turn out to be quite important later.

\section{Linking the UV to the IR}
\label{link}

In this section we shall analyze the UV limit carefully. We start
from eq. (\ref{paolo}), noticing  first that  \begin{equation}
Y(\rho)= 4\rho -1 +8\rho\,\sum_{n=1}^{\infty} e^{-4n\rho},
\end{equation} and hence the argument $k$ of the complete elliptic
integral becomes for large values of $\rho$ (i.e. in the UV)
\begin{equation} k^2=1-\frac{1}{Y(\rho)}\approx
1-\frac{1}{4\rho}+... ~. \end{equation} The quantity
${k^{\prime}}=\sqrt{1-k^2}$ becomes ${k^{\prime}}^2\approx
1/4\rho$. We hence need to expand around ${k^{\prime}}=0$, where
$E(k)$ is not analytic in ${k^{\prime}}$. Using the well known
expansion of $E$ near $k=1$
\begin{equation} E(k)\approx
1+\frac{1}{2}\left(\ln\frac{4}{{k^{\prime}}}-\frac{1}{2}\right){k^{\prime}}^2+
O({k^{\prime}}^4\ln {k^{\prime}}), \end{equation} we obtain from
these equations and eq. (\ref{paolo}) \begin{equation}
\frac{16\pi^2}{g_{\rm YM}^2 N}\approx 4\rho +\frac{1}{2}\ln
(8\sqrt{\rho}) -\frac{5}{4}+O\left(\frac{\ln \rho}{\rho}\right)+O(\rho
e^{-4\rho}). \label{g^2} \end{equation} This shows that
$1/g^2_{\rm YM}$ is not analytic in $\rho$ for large $\rho$.
{}From this result we can compute $\beta_\rho$ by differentiation,
\begin{equation} \beta_\rho (g_{\rm YM})\approx -\frac{g^3_{\rm
YM}N}{8\pi^2}\left(1 +\frac{1}{16\rho}+...\right). \end{equation}
The $\beta-$function referring to the scale $L=\ln (\mu/\Lambda)$
is:
\begin{equation} \beta(g_{\rm YM})\approx -\frac{g^3_{\rm
YM}N}{8\pi^2}\left(1
+\frac{1}{16\rho}+...\right)\frac{\partial\rho}{\partial L}.
\label{rightbeta} \end{equation} In passing we mention that eq.
(\ref{g^2}) can be inverted (iteratively) to give \begin{equation}
\rho\approx \frac{4\pi^2}{g^2_{\rm YM}
N}+\frac{1}{16}\ln\frac{g^2_{\rm YM}N} {256\pi^2}+\frac{5}{16}+...,
\label{invg^2}\end{equation} showing that $\rho$ is not analytic in
$1/g_{\rm YM}^2$ for a small YM coupling.

We now return to eq. (\ref{new}). Since $g_{\rm YM}$ is a function
of $\rho$ from eq. (\ref{paolo}), we can equally well write eq.
(\ref{new}) as \begin{equation} \frac{\Lambda^3}{\mu^3}=g(\rho
)~a(\rho). \label{zz} \end{equation} It should again be emphasized
that since the function $f$ in eq. (\ref{new}) is not unique, but
scheme dependent, the same applies to the related function $g(\rho
)$.

We now want to determine the unknown function
$g(\rho)$ from the following three requirements:
\begin{itemize}
\item[(i)]{The first two coefficients in the beta-function
(\ref{rightbeta}) should be correctly reproduced, since they are
universal \cite{Hooft-Book} for a wide range of renormalization
schemes.} \item[(ii)]{There should be an infrared limit,
corresponding to $\Lambda/\mu\rightarrow\infty$.} \item[(iii)]{
Point (i) and (ii) are achieved (hopefully) assuming the function
$g(\rho )$ to be a simple global factor.}
\end{itemize}

We now claim that the following version of eq. (\ref{zz}), which
satisfies (iii), \begin{equation}
\frac{\Lambda^3}{\mu^3}=\rho^{-p} ~a(\rho), \label{rhop}
\end{equation} with some power $p$ to be determined from the
requirement (i), also satisfies the two requirements (i) and (ii)
simultaneously. We notice that the case $p=0$ corresponds to the
choice in \cite{DiVecchia:2002ks}, so the calculations of these
authors are included in the following.

{}From (\ref{rhop}) and the asymptotic UV behavior
\begin{equation} a(\rho )\approx 4\rho e^{-2\rho}
~(1+e^{-4\rho}+...), \end{equation} we easily obtain
\begin{equation} \frac{\partial\rho}{\partial L}\approx
\frac{3}{2}\frac{1} {1-\frac{1-p}{2\rho}}\approx \frac{3}{2}
\left(1+\frac{1-p}{2\rho}\right). \end{equation} It should be
noticed that since we consider the large $\rho$ limit, the
expression $1/(1-(1-p)/2\rho)$ is not more accurate than
$1+(1-p)/2\rho$, since terms of order $1/\rho^2$ arises also from
other sources, for example the further expansion of the complete
elliptic integral.

The beta-function (\ref{rightbeta}) then becomes
\begin{eqnarray}
\beta (g_{\rm YM})&\approx & -\frac{g^3_{\rm YM}N}{8\pi^2}\left(1+
\frac{1}{16\rho}+...\right)\frac{3}{2}\left(1+
\frac{1-p}{2\rho}\right)\nonumber \\
&\approx &-\frac{3g^3_{\rm YM}N}{16\pi^2}\left[1+
\left(\frac{9}{8}-p\right)\frac{g^2_{\rm YM}N}{8\pi^2}+...\right].
\label{pbeta}
\end{eqnarray}
In the fifth order term there are three contributions, namely a
term \cite{error}  $+1/16\rho$ coming from the
expansion of the complete elliptic integral, a term $-p/2\rho$
coming from the factor $\rho^{-p}$ in eq. (\ref{rhop}), and a term
$+1/2\rho$ coming from the function $a(\rho )$.

{}It is also instructive to display the beta function containing
the next order in the expansion in $\frac{1}{\rho}$
\begin{eqnarray}\beta (g_{\rm YM})&\approx & -\frac{3 g^3_{\rm YM} N}{16\pi^2}\left[1+
\frac{1}{2\rho} \left(\frac{9}{8}-p\right) +
\frac{1}{4\rho^2}\times \right. \nonumber
\\ &\times& \left. \left( (1-p)\left(\frac{9}{8} - p\right) \right. \right. + \nonumber \\
&~& \left. \left. \frac{3}{4^3} \left(\frac{35}{12} - \ln
\left(8\sqrt{\rho}\right)\right)\right) \right] + \ldots\ ,
\end{eqnarray}
which as function of the coupling constant reads: \begin{eqnarray}
\beta (g_{\rm YM}) &\approx &-\frac{3g^3_{\rm
YM}N}{16\pi^2}\left[1+ \left(\frac{9}{8}-p\right)\frac{g^2_{\rm
YM}N}{8\pi^2} \right. \nonumber
\\&+&  2\frac{g^4_{\rm YM}N^2}{4^5 \pi^4} \Big[8\left(p^2-\frac{3}{2}~p+
\frac{143}{256}\right)\nonumber \\
&+& \left. \left.  \left(p - \frac{15}{16}\right)\ln
\left(\frac{g^2_{\rm YM}N}{256 \pi^2}\right)\right] \right] +
\ldots  \label{p2beta}
\end{eqnarray}
{}In showing this result, for completeness, we kept not only the
leading logarithmic corrections but we also display the constant
terms appearing in the coefficient of the order $g^7_{\rm YM}$
term. The  $g^7_{\rm YM}$ term in eq. (\ref{p2beta}) contains a
logarithmic term which does not occur in a straightforward perturbative
calculation of the beta function. However, by a coupling constant
transformation it is possible to remove the logarithmic term. Likewise
eq. (\ref{p2beta}) can be put in agreement
with the correspondent term in eq.~(\ref{betaym}) via a coupling
constant transformation.

To see how this work in some detail let's schematically write the beta
function we derived as:
\begin{eqnarray}
\beta(g) = c_{1}g^3 + c_{2} g^5 +g^7 (c_{3}+c_{4}\ln g) + \ldots \
,
\end{eqnarray}
with the obvious identification of the $c_{i}$ coefficients and
coupling constant $g$ with our previous expression. Let's perform
the following coupling constant change
\begin{eqnarray}
g = \tilde{g} + h_{1}\tilde{g}^3 + \tilde{g}^5\left(h_{2} + h_3 \ln
\tilde{g} \right) + \ldots \ .
\end{eqnarray}
In the new coupling constant the first two terms in the beta
function are unchanged (as expected by the universality argument)
while only the coefficient of the third term (i.e. the $g^7$ term)
changes:
\begin{eqnarray}
\tilde{\beta}({\tilde{g}})&=&c_{1}\tilde{g}^3 + c_{2}\tilde{g}^5 \nonumber
\\
&+& \tilde{g}^7 \left[ c_3 + 2c_2 h_1  \right. \nonumber \\ &+&
\left. c_1 (3 h_1^2 - 2 h_2 -h_3) + \left(c_4 - 2 c_1
h_3 \right)\ln \tilde{g} \right] \nonumber \\
\end{eqnarray}
Taking $h_3=c_4/2c_1$ we arrive at a beta function without the
logarithmic term. Clearly higher orders can always be fixed by a
suitable non universal redefinition of the coupling constant
\cite{Hooft-Book}.

At this point we can focus directly on the universal terms.
 {}From (\ref{pbeta}) we see that irrespective of the value of
$p$ we always obtain the leading asymptotically free term in the
beta function. However, the next term fixes the value of $p$,
since the first two universal terms should correspond to the beta
function \begin{equation} \beta (g_{\rm YM})\approx
-\frac{3g^3_{\rm YM}N}{16\pi^2}~\left(1+ \frac{Ng^2_{\rm
YM}}{8\pi^2}\right). \end{equation} Thus $9/8-p=1$, or
\begin{equation} p=\frac{1}{8}. \end{equation} This value of $p$ thus
ensures the validity of the condition (i).

At this stage it should be pointed out that if we had started with
an arbitrary function $g(\rho )$ in eq. (\ref{zz}), we would then
have found that in order to satisfy the condition (i) we would
need that $g(\rho )\approx \rho^{-1/8}$ to leading order when
$\rho\rightarrow\infty$. In this sense eq. (\ref{rhop}) is a
consequence of (i) in the UV. Promoting (\ref{rhop}) to a global
equation is, of course, a separate assumption.

To see what happens in the infrared, we notice that $a(\rho
)\rightarrow 1$ for $\rho\rightarrow 0$. Thus, from eq.
(\ref{rhop}) we see \begin{equation} \rho\approx
\left(\frac{\mu}{\Lambda}\right)^{\frac{3}{p}}. \end{equation}
This asymptotic behavior can be directly compared to the behavior
in eq. (\ref{important}), giving the result \begin{equation}
\beta_0=\frac{6}{p}=48, \label{result} \end{equation} valid for a
{\it first order} zero in the beta-function. We remark that higher
order zeros are not possible due to the requirement (iii), since
they require $\rho$ to behave like a power of $-L=-\ln
\mu/\Lambda$. Such a power could only arise if $\rho^{-p}$ in eq.
(\ref{rhop}) is replaced by \begin{equation} \rho^{-p}\rightarrow
\exp \left[ \frac{3(g_c\rho^2)^{(1-\alpha)}}{\beta_0(\alpha
-1)}\right], \end{equation} where $\alpha
>1$ is the order of the fixed point. In the IR the expression on the
right hand side behaves like $e^{-3L}$ due to the second eq. (\ref{important}).
In the UV this would lead to
\begin{equation} \frac{\partial\rho}{\partial L}\approx
\frac{3}{2}\left(1+\frac{1}{2\rho}
-\frac{3}{\beta_0}\frac{(g_c\rho^2)^{1-\alpha}}{\rho}+\dots
\right)~. \end{equation} To satisfy the requirement (i) we need
that
\begin{equation}
\frac{3}{\beta_0}g_c^{1-\alpha}\rho^{1-2\alpha}=\frac{p}{2\rho}+\ldots\
, \end{equation} which is impossible due to the condition $\alpha
>1$ for a higher than first order fixed point. Therefore we need
$\alpha =1$.

Consequently we see that the beta-function behaves like
\begin{equation} \beta (g_{\rm YM})\approx -48~ (g_c-g_{\rm
YM})\approx -\frac{6\sqrt{N}} {\sqrt{2\pi}}~(g_c^2-g^2_{\rm YM}).
\label{final} \end{equation} This result shows that the behavior
of beta near the IR zero has been fixed by requiring that the
second term in the UV beta function has the correct universal
value.

Our computations and results can be straightforwardly applied to
Ref.~\cite{Imeroni:2002me} which make use of a different
nonsingular supergravity solution based on the warped deformed
conifold found in \cite{Klebanov:2000hb}.

\section{Conclusions}
\label{The End}

We investigated the ${\cal N}=1$ four-dimensional super Yang-Mills
theory with gauge group $SU(N)$ using the approach introduced in
\cite{DiVecchia:2002ks}. Here the authors exploited the connection
between the gaugino condensate and the function $a(\rho)$. A
remarkable feature of this connection is that it leads to
asymptotic freedom with the correct coefficient. On the other hand
this relation allows more freedom which we used to investigate the
IR limit $\mu\rightarrow 0$ of the theory.

The supergravity theory provides a dimensionless scale $\rho$
respect to which we defined a beta function $\beta_\rho$. The
latter is of direct relevance for the super Yang-Mills theory when
making the mild assumption that $\rho$ is some function of the
field theory scale $\mu$. $\beta_\rho$ is completely determined
and displays an infrared fixed point. Assuming that this has a
physical meaning, the beta functions computed with respect to the
scale $\mu$ should then have this behavior as well. This still
leaves much freedom. However, if the relation between the gaugino
condensate and the function $a(\rho)$ contains a ``global'' (i.e.,
the same factor in the IR and the UV) factor, then the order of
the fixed point and the slope of the beta function in the IR can
be uniquely fixed using the UV information. The UV input is due to
the ``universality'' of the first two terms of the perturbative UV
beta function. Clearly this input is not intrinsic to the approach
presented in \cite{DiVecchia:2002ks}, but must be taken as a kind
of boundary condition from the known results regarding the super
Yang-Mills theory in the UV.

The existence of an IR fixed point in our approach leads to the
natural question concerning the physical meaning of this. For
example, one may ask if the four dimensional theory is still
confining. On this account it worth recalling that the
renormalization invariant scale $\Lambda$ is always there, and
survives also in the infrared, where the vanishing of $\mu$ is
exactly compensated by the zero in the beta function in the
construction of $\Lambda$. Hence one needs independent arguments
to settle the confinement question. For a recent discussion of
this question we refer to \cite{davies}.

The most natural interpretation of a Landau singularity is that
the original, weak coupling (semi-classical) vacuum is unstable (tachyonic),
or meaningless (ghost-like), and is replaced by
another ground state. Similarly, the existence of an infrared
fixed point can be interpreted as the persistence of the original
vacuum state. In the case of ${\cal N}=1$ super Yang-Mills theory
this can be related to the persistence of the gaugino condensate,
i.e. the expectation value \begin{equation}
<0|\lambda^2|0>=\Lambda^3. \label{vacuum} \end{equation} There is
no sign that the vacuum state $|0>$ needs to decay to another
state \cite{Shifman:1999kf},\cite{davies}. The constant $\Lambda$ is well defined in
the whole range of the scale $\mu$ from $\infty$ to 0. Therefore
the physical meaning of the IR fixed point may be that it leaves
the gaugino condensate as a genuine physical feature of the
theory. This has to be contrasted with the case where there is a
Landau singularity. The renormalization group invariant $\Lambda$,
originally defined in the weak coupling regime,
would then only be defined up to a scale $\mu =\Lambda$, beyond
which the theory with vacuum $|0>$ would be ill defined. In a new
vacuum, which hopefully exists beyond the Landau singularity, other
condensations than (\ref{vacuum}) may then be
preferred. We therefore think that the IR fixed point advocated by
us is a physically reasonable feature of ${\cal N}=1$ super
Yang-Mills theory, since it keeps the gaugino condensation unchanged at
large couplings.

 \acknowledgments
It is a pleasure to thank Poul Henrik Damgaard, Paolo di Vecchia,
Alberto Lerda and Raffaele Marotta for helpful discussions
and comments. The work of F.S. is supported by the Marie--Curie
fellowship under contract MCFI-2001-00181.

\appendix
\section{A Different Choice}
As repeately emphasized the choice of the interpolating function
$\rho^{-p}$ exhibited in (\ref{rhop}) is not unique. To illustrate
this point we now consider the Bessel function $I_\nu (z)$ with a
suitable relation between $z$ and $\rho$. The asymptotic behaviors
are:
\begin{eqnarray} \lim_{z\rightarrow \infty}I_\nu (z)&\approx& \frac{e^{z}}{\sqrt{2\pi
z}}+\ldots, \\  && \nonumber \\ \lim_{z\rightarrow 0}I_\nu
(z)&\approx& \frac{1}{\Gamma (\nu
+1)}\left(\frac{z}{2}\right)^\nu\left[ 1+\frac{1}{\nu
+1}\left(\frac{z}{2}\right)^2+ \dots\right]. \nonumber \\
\label{zto0}
\end{eqnarray}
Now in the IR we need an exponential behavior (which corresponds
to $z\rightarrow \infty$), so we need to set \begin{equation}
z=\frac{3~(g_c\rho^2 )^{1-\alpha }}{\beta_0 (\alpha -1)}.
\end{equation} Here we have $\alpha >1$.

On the other hand, in the UV (i.e. $z\rightarrow 0$) we need
\begin{equation} \rho^{-1/8}\propto \rho^{2\nu (1-\alpha)}, \end{equation}
leading to \begin{equation} \alpha=1+\frac{1}{16\nu },
\label{101}\end{equation} which is indeed larger than 1. Hence we
can have the following behavior, \begin{equation}
\frac{\Lambda^3}{\mu^3}=I_\nu \left(\frac{3~(g_c\rho^2 )^{1-\alpha
}} {\beta_0 (\alpha -1)}\right)~a(\rho ),
\label{105}\end{equation} and
\begin{equation} \beta(g_{\rm YM})=-\beta_0 (g_c-g_{\rm
YM})^{1+\frac{1}{16\nu}}. \end{equation} So according to this
result it seems that by a different choice of the function
$g(\rho)$ (still allowing a universal behavior in the UV) one can
have a higher order IR fixed point.

However, if we consider the asymptotic expansion (\ref{zto0}) we
see that the correction terms are of order $z^2$ relative to 1,
i.e. of order $\rho^{4(1-\alpha )}$. Since this is a perturbative
correction in the UV, and since $\rho\propto 1/g_{\rm YM}^2$, we
observe that this correction corresponds to $(g_{\rm
YM}^2)^{4(\alpha -1)}$. In order for this to be really
perturbative, we require \begin{equation} 4(\alpha -1)=1, \quad
{\rm i.e.}\quad\alpha=1.25. \label{100}\end{equation} Although
this power is larger than one, it is fractional.  This would lead
to a non analytical behavior of the beta function with undesirable
branch cuts not expected in field theory.

We mention that the result (\ref{100}) also results if one
uses eq. (\ref{105}) in the UV,
\begin{equation}
e^{-3L}\approx \rho^{2\nu (1-\alpha )}e^{-2\rho}\rho\left(1+{\rm const.}~
\rho^{4(1-\alpha )}\right),
\end{equation}
leading to
\begin{equation}
\frac{\partial\rho}{\partial L}\approx \frac{3}{2} \left(1+
\frac{1+2\nu (1-\alpha )}{2\rho}+{\rm const.}~\rho^{3-4\alpha}\right).
\end{equation}
Demanding that the third term on the right hand side of this equation is of
order $1/\rho^2$ again gives eq. (\ref{100}). We also see that in order
to have the right coefficient in the second term (1-1/8) we again
obtain eq. (\ref{101}).

We cannot completely exclude the possibility that by a suitable
choice of the interpolating function the IR fixed point turns into
a higher order one. However the previous computations show that
such a possibility is unlikely.

If we insist that there should be a first order zero in the beta
function we see from the work in the main text that this requires
a power behavior of the function $g(\rho )$ in the infrared,
$\rho\rightarrow 0$. Also, to satisfy the requirement (i) we
always need the power -1/8. Consequently, in this case there is
very little freedom in the choice of $g(\rho )$. Taking into
account that the functional dependence can anyhow be changed by
coupling constant transformations (even if ${\cal N}=1$ super
Yang-Mills theory were completely solved), it appears that the
simple choice $g(\rho )=\rho^{-1/8}$ is reasonable.

\end{document}